\documentclass[twocolumn,trackchanges]{aastex701}

\pdfoutput=1
\usepackage{comment}
\usepackage{graphics}
\usepackage{graphicx}
\usepackage{amsmath,amsfonts,amssymb,amsthm}
\usepackage[normalem]{ulem}
\usepackage{alltt}
\usepackage{tabularx,tabulary}
\usepackage[varg]{txfonts}
\usepackage{float}
\usepackage{dcolumn}
\usepackage[nolist,nohyperlinks]{acronym}
\usepackage{xspace}
\usepackage[english]{babel}
\usepackage[abs]{overpic}
\usepackage{pict2e}
\usepackage{subcaption}
\allowdisplaybreaks[1]
\usepackage[utf8]{inputenc}
\usepackage{bm}
\usepackage{stackengine}
\usepackage{boldline,multirow}
\usepackage{braket}

\usepackage{rotating}
\usepackage{booktabs}
\usepackage{longtable}
\usepackage{adjustbox}

\graphicspath{{pics/}}
\begin{document}
\title{Inferring host environment properties and gravitational-wave decay time from the eccentricity measurement of dynamically captured binaries}
\correspondingauthor{A.~Vincent Paul}
\email{vincenta@cmi.ac.in}

\author[0009-0006-4764-2811]{A.~Vincent Paul}
\email{vincenta@cmi.ac.in}
\affiliation{Chennai Mathematical Institute, Siruseri, 603103, India}

\author[0000-0002-5490-2558]{Parthapratim Mahapatra}
\email{parthapratim.mahapatra@cardiff.ac.uk}
\affiliation{School of Physics and Astronomy, Cardiff University, Cardiff, CF24 3AA, United Kingdom}

\author[0000-0001-8270-9512]{Marc Favata}
\email{favata@montclair.edu}
\affiliation{Department of Physics \& Astronomy, Montclair State University, 1 Normal Avenue, Montclair, New Jersey 07043, USA}

\author[0000-0002-6960-8538]{K. G. Arun}
\email{kgarun@cmi.ac.in}
\affiliation{Chennai Mathematical Institute, Siruseri, 603103, India}
\affiliation{Max Planck Institute for Gravitational Physics (Albert Einstein Institute), D-30167 Hannover, Germany}
\affiliation{Leibniz Universit\"at Hannover, D-30167 Hannover, Germany}

\date{\today}
\begin{abstract}
Dynamical capture in dense stellar environments is a promising channel for producing eccentric compact binary mergers. Although there have been no confident detections of eccentric mergers to date, a few candidates show indications of non-negligible in-band eccentricity upon re-analysis of the data. By assuming an observed eccentric event originates from a dynamical gravitational wave (GW) capture, we show that it is possible to identify the host environment using the eccentricity and mass posteriors. 
In particular, the eccentricity posterior can be mapped to posteriors on key capture parameters, such as the relative velocity at infinity and the impact parameter. By comparing these with the expected velocity distributions of different astrophysical environments, we can place constraints on the likely host. Assuming that it originated from a GW capture, we applied this framework to the neutron star-black hole merger GW200105. By comparing with the velocity dispersion distributions of neutron stars in the cores of globular clusters (GCs) and nuclear star clusters (NSCs), we find the probability that GW200105 merged in a GC (NSC) to be $\sim 29\% \,(71\%)$. As we anticipate detecting several eccentric mergers in the future, this method can provide a valuable astrophysical diagnostic of their host environments on a single-event basis; this can be straightforwardly generalized to a population of eccentric binaries. The formalism we develop is also applied to GW190521, but is less constraining for that event. Lastly, we infer a GW decay time from capture to merger of $11\mbox{--}156$ days for GW200105.
\end{abstract}

\section{Introduction}

Since the landmark discovery of the binary black hole merger, GW150914~\citep{AbbottGW150914}, the LIGO, Virgo, and KAGRA (LVK) observatories has observed hundreds of compact binary coalescences, including binary black holes (BBH), binary neutron stars (BNS), and neutron star-black hole (NSBH) systems \citep{GW170817,GWTC-1-pop,GWTC1catalogue,AbbottGW200105,GWTC-3,GWTC2.1,Abac_2024GW230529,Abac_2025_GWTC4}. A sought-after feature in these signals is orbital eccentricity, which carries information about a compact binary's formation history. 
Studies have shown that the minimum eccentricity measurable by Advanced LIGO~\citep{Aasi_2015} is approximately $e_{10} \sim 0.05\mbox{--}0.1$
for typical signal-to-noise ratios~\citep{Brown_zimmerman_ecc_detectability, Lower2018, RomeroShaw2019,Favataecc2021, Saini_2024}. Here, $e_f$ is the eccentricity at a reference frequency $f$. Although the LVK collaboration has not yet reported a confident detection of eccentricity, recent studies~\citep{RomeroShaw2020GW190521,Gupte:2024jfe,Gupta_eccentric_subpopulation,KacanjaEcc} suggest that a subpopulation of eccentric binaries may already be present in existing Gravitational-Wave (GW) catalogs. Those events were previously interpreted as quasicircular because the waveform models did not include eccentricity. A notable example is GW200105\_162426 (hereafter GW200105), the first NSBH merger confidently detected by LIGO Livingston and Virgo~\citep{AbbottGW200105}. This event has source-frame masses $m_{\rm NS} =1.9 M_{\odot}$ and $m_{\rm BH}=8.9 M_{\odot}$. Several independent analyses show it to be consistent with an eccentricity of $e_{20}\sim 0.1$  ~\citep{MorrasEccWaveform,Morras2025,Planas:2025plq,Aasim2025,Tiwari2025GW200105Eccentricity,KacanjaEcc,BhatECCtest,planas200208andothers,Phukon2026,Morras2026}. Another example is the high-mass BBH merger GW190521~\citep{GW190521,GW190521_apjl}  with source-frame masses $\sim 85 M_{\odot}$ and $66 M_{\odot}$. Subsequent analyses employing eccentric waveforms suggest GW190521 could be consistent with a significantly eccentric binary with  $e_{10} \gtrsim 0.1$ ~\citep{Gayathri2020NR,RomeroShaw2020GW190521}. However, the eccentricities inferred by different models differ considerably, likely due to the short signal duration. Also, more recent estimates suggest that GW190521 does not show prominent signs of eccentricity~\citep{RamosBuades2023,Gupte:2024jfe,Planas_2026}. This makes any meaningful astrophysical interpretation difficult. Other GW events have also been suggested as potential eccentric candidates, including GW200208\_222617~\citep{Romeroshaw200208,planas200208andothers} which shows less ambiguous  signs of eccentricity. Events such as GW200129, GW190701, and GW190929~\citep{planas200208andothers,Gupte:2024jfe} also show signs of eccentricity; however, the evidence in these cases remains inconclusive.  Nevertheless, the inference of relatively large eccentricity in a GW event suggests that it is unlikely to form via isolated binary evolution (see \citet{BenacquistaLiving, Mandel:2021smh} for reviews).
Residual eccentricity in the LVK band provides a powerful diagnostic of compact binary formation channels ~\citep{Samsing2014EccentricEncounters,samsing2018eccmergers,Zevin2021,Cheng2023,Isobel_res_ecc_2024}. Dynamically formed binaries can retain measurable eccentricity when entering the LIGO frequency band~\citep{Tremaine_2017_isolatedtriples,samsing_ecc_GC_bin_sin,samsing2018eccmergers,Samsing_sin_sin_GWcap2019,Samsing2014EccentricEncounters}. Below, we summarize two principal channels proposed to produce eccentric binary mergers.

Hierarchical triple systems provide an efficient channel for eccentric compact binary mergers~\citep{Shariat_2024_triple,Attia2025}. Perturbations from a tertiary companion induce Kozai--Lidov oscillations~\citep{KozaiOsc62,naoz_triples_2013,Kimpson_2016,Naoz_2017,Tremaine_2017_isolatedtriples,gupta_triples,Trani_2022}, driving the inner binary to extreme eccentricities and enabling measurable eccentricity in the detector band. Population studies suggest that a significant fraction of eccentric mergers may be produced by this channel~\citep{Wen2003,Aarseth:2012MNRAS,Antonini_2016_triples,Antonini2017FieldTriples,Morras2026}, with recent simulations~\citep{Stegmann2025} showing that rates of NSBH systems from triples can match the inferred local merger rate from LVK observations.

Dynamical interactions in dense stellar environments such as globular clusters (GCs) ~\citep{Morscher_2015_dyn_GC,Hurley_Sippel_Tout_Aarseth_2016,CBC_astrophysics_review,Kritos_2024}, young star clusters~\citep{Banerjee_2017,DiCarlo_2020,Trani_2022}, and nuclear star clusters  (NSCs)~\citep{Miller:2008yw,Chattopadhyay:2023pil} provide another key pathway for eccentric mergers. Frequent close encounters, including single-single~\citep{Bae_2017_sin_sin,Samsing_sin_sin_GWcap2019,Hoang_2020}, binary-single ~\citep{Samsing_2017,samsing_ecc_GC_bin_sin,Gamba2023GW190521,Rando_bin_sin}, and binary-binary interactions ~\citep{Zevin_2019_bin_bin}, can efficiently form and harden compact binaries, allowing them to merge in a Hubble time. If two unbound compact objects pass sufficiently close, the radiated GW energy can exceed the kinetic energy, leading to a bound, highly-eccentric binary that rapidly merges while retaining significant eccentricity in the frequency band of  ground-based detectors~\citep{Hansen74,Hills1980,Quinlin_shapiro_87,quinlin89,OLeary2009,OLeary2016ApJ,Bae_2017_sin_sin,cho2018_GW_hyperbolicorbit,Gondan2018GWCaptureNuclei,Samsing_sin_sin_GWcap2019,Gondan2020capGN}. This is referred to as  GW capture. 
In addition, binary-single interactions can excite large eccentricities during resonant encounters~\citep{Samsing2014EccentricEncounters,Leigh2018DynamicalHardening}, producing mergers with $e_{10} \gtrsim 0.1$~\citep{samsing_ecc_GC_bin_sin}.  As both hierarchical triples and dynamical captures can produce binaries that retain measurable eccentricity at detection, eccentric GW signals provide a powerful probe of the astrophysical origin of compact binary mergers.  Based on general considerations, \cite{Rozner2026} argued that the eccentricity distributions expected from various dynamical formation channels are likely to be very similar at the frequencies probed by LVK detectors.
 
Recently,~\citet{Gupte2026} reanalyzed the BBH events in the first part of the LVK's fourth observing run (O4a) using a multipolar, eccentric, spin-aligned, effective-one-body waveform model~\citep{SEOBNRv5PHM2023,SEOBNRV5EHM2025}. They found no events with significant evidence for eccentricity. Assuming these binaries formed via single-single scattering (GW capture), \citet{Gupte2026} investigated the type of star cluster in which these could have occurred. Using a Maxwellian distribution as a reference model for the relative velocity distribution of compact objects, they constrained the velocity dispersion parameter $\sigma$ using the non-detection of eccentricity, concluding that NSCs can be ruled out as potential hosts of these mergers. 

Here, we address this question in the specific context of GW capture, but using single events with informative eccentricity posteriors. This complements the approach of \citet{Gupte2026} which used eccentricity upper-limits from a population of events. Using relations between the unbound and capture orbital parameters, combined with a post-Newtonian (PN) evolution of the binary into the detector band, we develop a hierarchical Bayesian inference method  that constrains the host environment parameters given single-event mass and eccentricity posteriors.

\section{\label{sec:ecc-scatter}Relating the eccentricity posterior to the scattering parameters}
GW capture of compact objects can be modeled as a two-body scattering problem with GW emission providing dissipation. Two key parameters determining the capture criteria are the relative velocity of the scatterers at infinity $v_\infty$ and the impact parameter $b$. Consider two compact objects with masses $m_1>m_2$, total mass $M$, and symmetric mass ratio $\eta={m_1 m_2}/{M^2}$; they approach each other with an asymptotic relative velocity $v_\infty$. For two-body Newtonian scattering in the non-relativistic ($v_\infty\ll1$) and strong focusing  limit ($bv_\infty^2\ll M$, corresponding to the pinhole regime in \citet{Rozner2026} for successful GW capture), the orbit can be assumed to be nearly parabolic with eccentricity
\begin{equation}
e=\sqrt{1+\left(\frac{b\,v_\infty^2}{M}\right)^2}\approx 1.
\end{equation}
In this near-parabolic limit, the radiated GW energy at periapsis can be approximated as \citep{Hansen1972energy_radiated,Turner1977,Quinlin_shapiro_87,OLeary2009}
\begin{equation}
    \Delta E_{\rm GW}(r_p) \approx 
    \frac{85 \pi \eta^2 M^{9/2}}
    {12 \sqrt{2}\, r_p^{7/2}},
    \label{energylostGW}
\end{equation}
where $r_p$ denotes the periapsis distance. (Throughout, we use units where $G=c=1$.) Neglecting any perturbations outside the binary, the GW energy loss transitions the unbound binary to a bound one with $e\lesssim 1$. 
     
For a given $(b,v_\infty)$, we can express the periapsis distance and initial semi-major axis of the newly formed binary as~\citep{OLeary2009}
\begin{eqnarray}
\label{eq:rp}
r_p &=&\left(\sqrt{\frac{1}{b^2} + 
        \frac{M^2}{b^4 v_{\infty}^4}}
        + \frac{M}{b^2\,v_{\infty}^2}
    \right)^{-1} \; , \\
    a_{i} &=&
    \frac{M^2 \eta}
    {2\left| \frac{1}{2} \mu v_{\infty}^2 -\Delta E_{\rm GW}  
    \right|}.\label{eq:aini}
\end{eqnarray}
The corresponding eccentricity immediately after capture is   $e_i = 1-r_p/a_i$. 
For a fixed value of $v_\infty$, the maximum impact parameter for which a GW capture occurs can be obtained from the above  \citep{OLeary2009}:
\begin{equation}
    b_{\max} =
\left(\frac{340\pi}{3}\right)^{1/7}
\frac{\eta^{1/7} M}{v_{\infty}^{9/7}} \label{eq:bmax}.
\end{equation}

Expressing the semi-latus rectum $p = a(1-e^2)$ in a dimensionless form as $x \equiv p/M$, the PN orbital evolution due to GW emission from an initial bound binary configuration $(x_i,e_i)$ to a later configuration $(x_f,e_f)$  is given by~\citep{TuckerandWill2021}
\begin{align}
\label{eq:x_e}
x_f = x_i
\left(\frac{1+2/x_i}{1+2/x_f}\right)
\left(\frac{1-4/x_i}{1-4/x_f}\right)^{12/19}
\frac{g(e_f)}{g(e_i)} \,,
\end{align}
where $g(e)=e^{12/19}(304+121e^2)^{870/2299}$~\citep{Peters1964}. Here, we adopt  twice the orbital frequency as the definition of our reference frequency, $f=\frac{1}{\pi}\sqrt{\frac{M}{a^3}}$. Combining Eqs.~\eqref{energylostGW}, \eqref{eq:rp}, \eqref{eq:aini}, and \eqref{eq:x_e}, we have a numerical prescription $e_f=F(b,v_{\infty},M,\eta)$ to compute the detector-band eccentricity $e_f$ given the binary's parameters prior to capture.
\begin{figure*}[ht]
\centering
\includegraphics[width=0.8\linewidth]{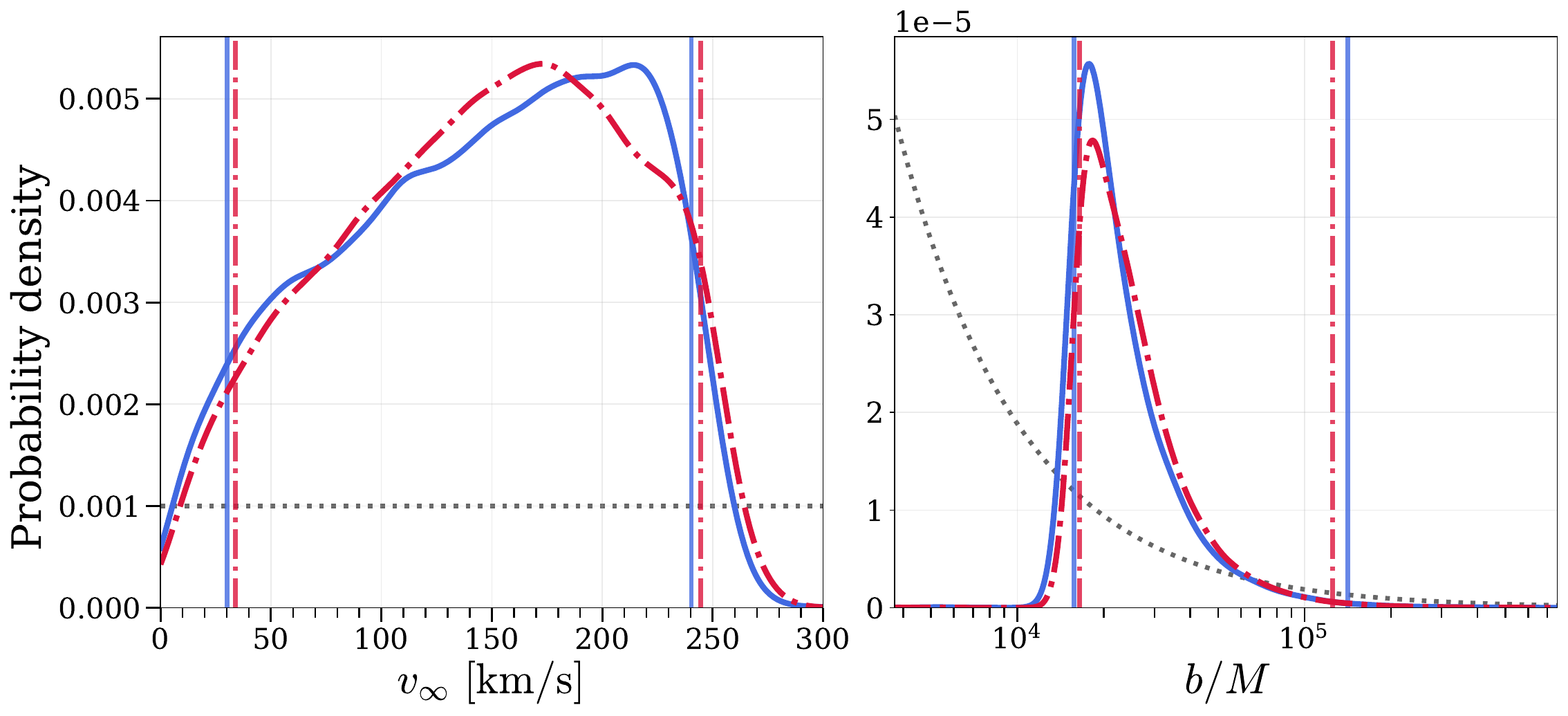}
  \caption{Inferred posteriors for $v_\infty$ and $b$ for GW200105 based on the formalism in Sec.~\ref{sec:bayesian}. As input we have used eccentricity posteriors for $e_f$ from two different analyses on GW200105 by \citet{Morras2025} (blue/solid) and \citet{KacanjaEcc} (red/dash-dotted). Dotted lines show the assumed prior distributions. Vertical lines show symmetric $90\%$ credible intervals.}   
  \label{fig:combined posteriors}
\end{figure*}

\section{Bayesian formulation of the problem}\label{sec:bayesian}
\subsection{\label{subsec:formalism}Hierarchical Bayesian inference of capture parameters}
Next, we develop a hierarchical Bayesian formalism   to infer the capture parameters $(b,v_\infty)$ given an eccentricity posterior $p(e_f|d)$ at a reference frequency $f$. Our approach here closely follows ~\citet{Mahapatra_2024,Mahapatra_2025}, though the context is totally different.
Let $\vec{\theta}=(b,v_\infty,M,\eta)$ denote the pre-capture parameters. 
Using Bayes' theorem, the posterior distribution of $\vec\theta$ is given by 
\begin{equation}
    p(\vec{\theta}|d)
\propto \Pi(\vec{\theta})
\mathcal{L}(d|\vec{\theta}),
\label{bayes theorem}
\end{equation}
where $\Pi(\vec{\theta})$ is the prior and $\mathcal{L}(d|\vec{\theta})$ is the likelihood.

The likelihood can be re-expressed as
\begin{equation}
   \mathcal{L}(d|\vec{\theta})=\int \mathcal{L}(d|e_f) p(e_f|\vec{\theta}) \, de_f ,,
   \label{Likelihood marginalization}
\end{equation}
 where $p(e_f|\vec\theta)$ is the probability of obtaining a particular eccentricity at some reference frequency $f$ given $\vec{\theta}=(b,v_\infty,M,\eta)$. Further, using Bayes' theorem, the likelihood $\mathcal{L}(d|e_f)$ can be expressed as
\begin{equation}
    \mathcal{L}(d|e_f) \propto \frac{p(e_f|d)}{\Pi(e_f)},
    \label{likelihood weightage}
\end{equation}
where $\Pi(e_f)$ is the eccentricity prior from the GW data analysis. 
From the mapping relations developed in the previous section, we have $e_f = F(b,v_\infty,M,\eta)$. Therefore, the quantity $p(e_f|\vec{\theta})$ in Eq.~\eqref{Likelihood marginalization} is a Dirac delta function that peaks at the value of $e_f$ determined by $\vec\theta$, implying
\begin{equation}
    p(e_f|\vec{\theta})= \delta\big(e_f - F(\vec{\theta})\big).
    \label{delta function probability}
\end{equation}
Hence, by substituting Eqs.~\eqref{likelihood weightage} and \eqref{delta function probability} into Eq.~\eqref{Likelihood marginalization}, Eq.~\eqref{bayes theorem} becomes

\begin{align}
    p(\vec{\theta}|d) &\propto \Pi(\vec{\theta}) \int \frac{p(e_f|d)}{\Pi(e_f)} \delta\big(e_f - F(\vec{\theta})\big) \, de_f , \, \nonumber \\
&\propto 
\Pi(\vec{\theta}) \frac{p(e_f|d)}{\Pi(e_f)}
\bigg\rvert_{e_f= F(\vec{\theta})}.
\label{posterior_general}
\end{align}

In practice, Eq. \eqref{posterior_general} is implemented as follows. First, we draw random samples from the prior distribution of $\vec \theta$ and apply the procedure in Sec.~\ref{sec:ecc-scatter} to obtain $e_f$. Then the likelihood for each sample is calculated by evaluating Eq.~\eqref{posterior_general}, where $p(e_f|d)$ and $\Pi(e_f)$ are provided by the parameter estimation analysis performed in \citet{Morras2025} and \citet{KacanjaEcc}; $\Pi(\vec{\theta})$ is discussed below. To generate discrete samples for the probability distribution function $p(\vec{\theta}|d)$, we use the {\tt emcee} sampler~\citep{emcee2013PASP}, which employs a Markov chain Monte Carlo ensemble sampling algorithm~\citep{Goodman2010CAMCS}.
In order to obtain a posterior on ($v_\infty, b$), we marginalize the posterior probability distribution in Eq.~\eqref{posterior_general} over $M$ and $\eta$:
\begin{equation}
    p(v_\infty,b | d) = \int p(\vec{\theta}|d)\, dM \,d\eta \,.
    \label{final posteriors}
\end{equation}
Equation \eqref{final posteriors} therefore  allows us to compute the posterior distribution of $(b,v_\infty)$ using a posterior distribution on $e_f$.

\subsection{Choice of Priors} 
For $v_\infty$, the prior distribution strongly depends on the binary's dynamical environment. A standard practice is to assume a Maxwellian velocity distribution for a specific stellar cluster (see, e.g., \citet{Lutzgendorf2012_maxw}). However, we avoid assuming a particular cluster model  at this stage and instead adopt simple, environment-agnostic priors. We assume a uniform prior between $v_{\infty}^{\rm min}$ and $v_{\infty}^{\rm max}$ for the relative velocity, 
 \begin{align}
 \Pi(v_\infty) = \frac{1}{v_\infty^{\rm max} -v_\infty^{\rm min}} \, ;\,(v_\infty^{\rm min},v_\infty^{\rm max})= (1,1000)\, {\rm km/s} \label{prior for v} \,.
\end{align}
 This spans different types of dense environments reasonably well.
For $b$ we adopt a prior motivated by the geometry of two-body scattering: $\Pi(b)\propto b$. This corresponds to sampling uniformly in the scattering cross-sectional area. Since GW capture is only possible for impact parameters smaller than the maximum capture value $b_{\max}(v_\infty,M,\eta)$ [Eq.~\eqref{eq:bmax}], the prior for $b$ must be normalized over the interval $b_{\rm min}<b<b_{\max}(v_\infty,M,\eta)$, where ~$b_{\rm min}=\frac{4M}{v_\infty}$ (corresponding to direct collision)   \citep{Gondan2018GWCaptureNuclei}. The normalized conditional prior therefore takes the form
\begin{equation}\label{eq:priorb}
\Pi(b|v_\infty,M,\eta)=\frac{2b}{b_{\max}^2-b_{\min}^2}\,.
\end{equation}
The joint prior for the capture parameters can then be written as
\begin{equation}
\Pi(b,v_\infty, M,\eta)=\Pi(b|v_\infty,M,\eta)\, \Pi(v_\infty, M, \eta)\,.
\end{equation}
 We further factorize $\Pi(v_\infty, M, \eta)=\Pi(v_\infty)\, \Pi(M, \eta)$. The uniform distribution in Eq.~(\ref{prior for v}) is used for $\Pi(v_{\infty})$. For the priors of $M$ and $\eta$, we use their posterior distributions as reported in the papers providing $p(e_f|d)$ \citep{Morras2025,KacanjaEcc}.

\subsection{\label{subsec:application}Application to GW200105}
We now apply our method to GW200105, which has been argued to have non-zero eccentricity by various independent studies.  
We use the publicly available data from \citet{Morras2025} and \citet{KacanjaEcc} for  the mass and eccentricity  posteriors. We then apply our Bayesian inference framework to compute the posteriors of the capture parameters $b$ and $v_{\infty}$. 

The results are shown in the top row of Figure~\ref{fig:combined posteriors}. At 90\% credibility, we obtain $v_\infty=151^{+89}_{-121}$ km/s and $b=2.54^{+11.56}_{-0.96} \times 10^4\,M$  using the posteriors in \citet{Morras2025}.  Using the posteriors in \citet{KacanjaEcc}, we obtain $v_\infty=151^{+93}_{-117}$ km/s and $b=2.75^{+9.75}_{-1.10} \times 10^4\,M$. These are in good agreement with each other, as shown in Figure~\ref{fig:combined posteriors}.

\section{\label{subsec:sigma_single_event}Inferring the host velocity dispersion from a single event}
Although a single GW capture event does not directly measure the velocity dispersion of its host environment, it can still provide information about the class of environments capable of producing such an encounter. The key idea is to use the event-level posterior of the relative velocity at infinity, $p(v_\infty|d)$ (Fig.~\ref{fig:combined posteriors}), to place constraints on the host-environment velocity dispersion through an environment-dependent capture model. To do this, we assume that the majority of a cluster's black holes and neutron stars have sunk to the cluster's center. We idealize all of those BHs and NSs to have the same mass, $m_{\rm BH}$ or $m_{\rm NS}$, respectively. The cluster center also contains tracer stars, all assumed to have a mass $m_{\star}$. We assume these subpopulations are each described by a velocity dispersion $\sigma_{i}$ (where $i=$ BH, NS, or $\star$) and that these subpopulations are in thermal equilibrium with each other. The equipartition theorem then implies that $m_i \sigma_{i}^2$ is a constant that is the same across all three subpopulations. Each subpopulation also follows a Maxwell-Boltzmann velocity distribution parameterized by $\sigma_i$.  

\begin{figure}[t]
    \centering
    \includegraphics[width=\columnwidth]{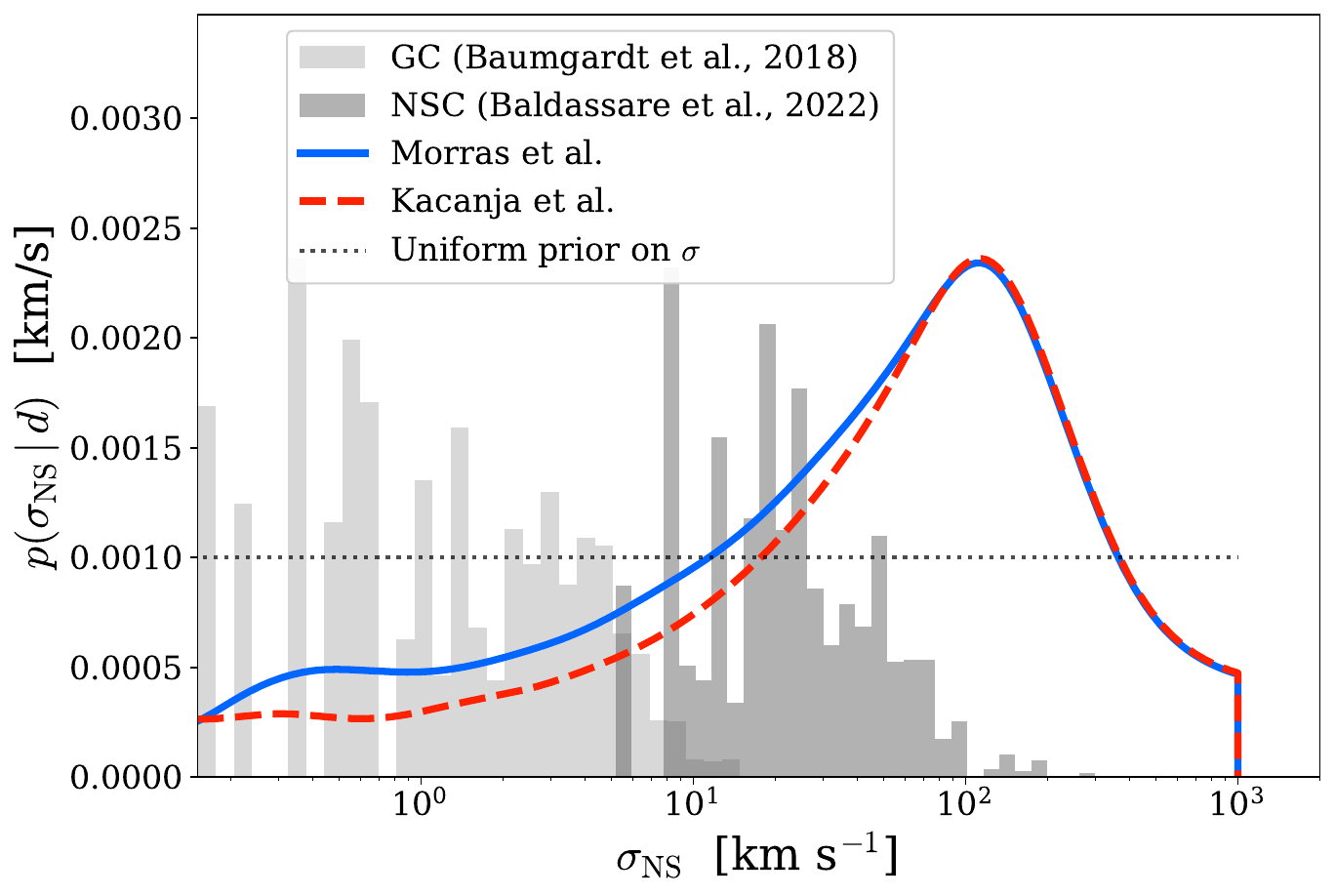}
    \caption{\label{fig:hosting probability 200105} Probability distribution of $\sigma_{NS}$ for the host environment of GW200105 obtained using the formalism in Sec.~\ref{subsec:sigma_single_event} with $\sigma_i=\sigma_{\rm NS}$ (red and blue curves). Gray histograms show observational constraints on $\sigma_{NS}$ based on rescaling stellar velocity dispersions via the equipartition theorem (see text for details). The relative heights of the GC and NSC distributions are also rescaled for better visualization.} 
    \end{figure}

Observationally, we can use the tracer star population to model (under the above assumptions) the compact object population of the cluster core. To do this, we make use of data in Table 2 of \citet{Baumgardt_velocity_disp_GC} for GCs and Table 1 of \citet{Baldassare2022MassiveBHNSC} for NSCs. For each environment (${\rm env}={\rm GC,\,NSC}$) we extract data for the stellar velocity distributions $p_{\rm env}(\sigma_{\star})$. These are rebinned to model the NS or BH velocity dispersion using $\sigma_{\star} = \sigma_i (m_i/m_{\star})^{1/2}$, where we use $m_{\star} =0.8 M_{\odot}$ and $m_{\rm NS}=1.4 M_{\odot}$ for the GW200105 analysis. These histograms are shown in gray in Figs.~\ref{fig:hosting probability 200105} and \ref{fig:sigmaposteriors-GW190521}. We fit a log-normal distribution to these histograms (see Table \ref{tab:lognormal_fits} in Appendix \ref{app:GW190521}); those fits are used for $p_{\rm env}(\sigma_i)$ in the analysis below. 

The distribution function for the relative 2-body velocity at infinity between two members of the cluster center is modeled as
\begin{align}
p(v_\infty|\sigma_i) \propto
v_\infty^2 \exp\!\left(-\frac{v_\infty^2}{2\sigma_{\rm rel}^2}\right)\,,
\label{Maxwell_dist}
\end{align}
where the relative velocity dispersion between objects of subpopulations $i$ and $j$ is given by $\sigma_{\rm rel}^2 = \sigma_{i}^2 + \sigma_{j}^2 = \sigma_{i}^2 (1+m_{i}/m_{j})$. If the subpopulations are similar in mass (i.e., $i=j={\rm BH}$ for a BBH encounter), then $\sigma_{\rm rel}^2 \approx 2 \sigma_i^2$. If $m_i \ll m_j$ (i.e., $i={\rm NS}$, $j={\rm BH}$ for a NSBH encounter), then $\sigma_{\rm rel}^2 \approx \sigma_i^2$. Hence, we assume the $v_{\infty}$ distribution can be characterized by a single velocity dispersion parameter $\sigma_i$. 
Because not all draws from the ambient relative velocity distribution lead to a GW capture, we must correct this distribution when studying GW capture binaries. The corresponding capture-weighted velocity distribution is therefore \citep{Gupte2026} 
\begin{align}
p_{\rm cap}(v_\infty|\sigma_i)
\propto
p(v_\infty|\sigma_i)\,
\sigma_{\rm cap}(v_\infty)\,
v_\infty,
\end{align}
where $\sigma_{\rm cap}(v_\infty)$ is the GW capture cross section and the extra factor of $v_\infty$ arises from encounter flux~\citep{Quinlin_shapiro_87,quinlin89}. For GW capture, the leading-order scaling of the capture cross section is  $\sigma_{\rm cap}(v_\infty) \propto v_\infty^{-18/7}$ \citep{quinlin89}.
Given the posterior $p(v_\infty|d)$ obtained from the single-event analysis, the posterior on the host velocity dispersion is constructed as
\begin{align}
p_{\rm cap}(\sigma_i | d)
\propto
\Pi(\sigma_i)
\int_{v^{\rm min}_\infty}^{v^{\rm max}_\infty}
\frac{p(v_\infty|d)}{\Pi(v_\infty)}
\,p_{\rm cap}(v_\infty|\sigma_i)\,dv_\infty,
\label{eq:sigma_posterior_general}
\end{align}
where $p(v_\infty|d)$ is given by Eq.~\eqref{final posteriors} marginalized over $b$, $\Pi(\sigma_i)$ is the prior on the velocity dispersion of the host's subpopulation and is chosen as $\Pi(\sigma_i) =\frac{1}{\sigma_{\rm max}-\sigma_{\rm min}}$ with $(\sigma_{\rm min},\sigma_{\rm max})= (1,1000)\, {\rm km/s}$, and $\Pi(v_{\infty})$ is the uniform prior in Eq.~\eqref{prior for v}.

Next, we assign a probablity $P({\rm env}|d)$ for each cluster type to host a particular GW capture event using $p_{\rm cap}(\sigma_i|d)$ via
\begin{align}
    P({\rm env}|d)=\frac{\int p_{\rm cap}(\sigma_i|d)\,p_{{\rm env}}(\sigma_i)\,d\sigma_i}{\int p_{\rm cap}(\sigma_i|d)\,[p_{\rm GC}(\sigma_i) + p_{\rm NSC}(\sigma_i)]\,d\sigma_i} \,,
    \label{prob cluster}
\end{align}
where $p_{{\rm env}}(\sigma_i)$ is the observationally motivated log-normal fit for a subpopulation's velocity dispersion.

We apply the above approach to estimate the velocity dispersion of GW200105. (GW190521 is also considered in Appendix \ref{app:GW190521}.) Figure \ref{fig:hosting probability 200105} shows the inferred posterior on $\sigma_{\rm NS}$ for the eccentricity posteriors from \citet{KacanjaEcc} and \citet{Morras2025}. Assuming the GW capture scenario, we see that GCs have $24\%\,(29\%)$ probability to host GW200105, while NSCs have a $76\% \,(71\%)$ probability using the eccentricity posteriors of \citet{KacanjaEcc} (or \citet{Morras2025}). This suggests that GCs are the  less likely host environment for GW200105. 
In Appendix \ref{app:GW190521} we applied this method to GW190521. In that case the results are less constraining; this is expected given the differences in the eccentricity estimates  between \citet{RomeroShaw2020GW190521} and \citet{Gayathri2020NR}. 

Our method also allows the computation of the decay time of a binary from the time of formation until it reaches the reference frequency in the detector's band. We find that the time for GW200105 to merge from the moment of capture and binary formation is between 11 and 156 days (with 90\% credibility). This is briefly discussed in Appendix \ref{app:decaytime}.

\section{Conclusions and Discussion}\label{sec:concl}
For compact binaries formed via dynamical capture in dense stellar environments, we proposed a method to relate the GW-inferred eccentricity to the relative velocity and impact parameter of the encounter. These quantities were then used to constrain the host environment. The method applies Newtonian capture conditions and a post-Newtonian evolution of the bound binary within a hierarchical Bayesian framework. Applying this to the possibly eccentric NSBH merger GW200105, we infer that NSCs are more probable to host GW200105 through a GW capture compared to GCs. The methodology developed here could serve as a valuable diagnostic tool for constraining the host environments of dynamically captured binaries in an era rich in events with non-negligible and precisely measured eccentricity. 

We acknowledge that the Keplerian eccentricity used here may diverge from waveform-based definitions in the strong PN regime; excluding spins in our evolution may also affect the estimation of the LIGO-band eccentricity. While these effects are unlikely to qualitatively change our conclusions, more precise modeling of the binary evolution would be desirable.  We also acknowledge that the assumption of full equipartition may not be achieved in real clusters, as the black holes at the cluster center may form their own gravitating sub-system. Lastly, these conclusions are made under the assumption that the binary is formed by GW capture. Any other mechanism that can produce this high eccentricity, like Kozai-Lidov oscillations in a hierarchical triple, can impact our inferences. It is, therefore, important to devise methods that can distinguish subpopulations of eccentric compact binaries.

\section{Acknowledgments}
We thank Anuradha Gupta for a review of this manuscript. A.V.P. and K.G.A.~thank Harald Pfeiffer, Nathan Johnson-McDaniel, B.~S.~Sathyaprakash and Anuradha Gupta for useful discussions. The authors thank Isobel Romero-Shaw for providing eccentric PE samples for GW190521. We also thank Isobel Romero-Shaw, Fabio Antonini, Aditya Vijayakumar and Juan Calder\'on Bustillo for several comments on the work. K.G.A.~is supported by Advance Research Grant ANRF/ARG/2025/000931/PS of the Anusandhan National Research Foundation and by the Max Planck Society. K.G.A.~and A.V.P. also acknowledge support from the Infosys Foundation. P.M.~acknowledges the Science and Technology Facilities Council for support through Grants ST/V005618/1 and UKRI2489. M.~F.~was supported by National Science Foundation (NSF) Grant No.~PHY-165337, and in part by the Heising-Simons Foundation and NSF grant PHY-2309135 to the Kavli Institute for Theoretical Physics.

This research has made use of data or software obtained from the Gravitational Wave Open Science Center (https://www.gwosc.org), a service of the LIGO Scientific Collaboration, the Virgo Collaboration, and KAGRA. This material is based upon work supported by NSF's LIGO Laboratory which is a major facility fully funded by the National Science Foundation. This document has LIGO preprint number {\tt LIGO-P2600185}.

\appendix
\section{Application to GW190521}\label{app:GW190521}
GW190521~\citep{GW190521} is a particularly remarkable BBH event, with component masses lying in the putative pulsational pair-instability supernova (PPISN) mass gap~\citep{PPSN_massgap_1967}. Following the LVK Collaboration's official data release, that event was re-analyzed using eccentric waveform models and found to be consistent with an eccentric binary merger~\citep{Gayathri2020NR,RomeroShaw2020GW190521}. However, the measured eccentricities of those two studies exhibit some discrepancy; this may be attributable to the short signal duration and differences in waveform models and priors.

We applied our method to the two eccentricity estimates for GW190521: (i) a moderately eccentric configuration with $e_{10} \sim 0.1$ \citep{RomeroShaw2020GW190521}, and (ii) a highly eccentric configuration with $e_{10} \sim 0.7$ \citep{Gayathri2020NR}. As the underlying data from \citet{Gayathri2020NR} are not publicly available, we construct an approximate eccentricity distribution that reproduces their median and  90\% credible interval.  We did not fit the mass posteriors; rather, we assume the median mass values from their analysis. This removes the marginalization step in Eq.~\eqref{final posteriors}.  We obtained the posteriors of  $\vec \theta=(b,v_\infty)$ following the method outlined in Sec.~\ref{sec:bayesian}. In the case of \citet{RomeroShaw2020GW190521}, we used their posteriors on $e_{10}$ and the masses to obtain the posteriors for $v_{\infty}$ and $b$. 

\begin{figure*}[t]
\centering
\includegraphics[width=0.8\linewidth]
{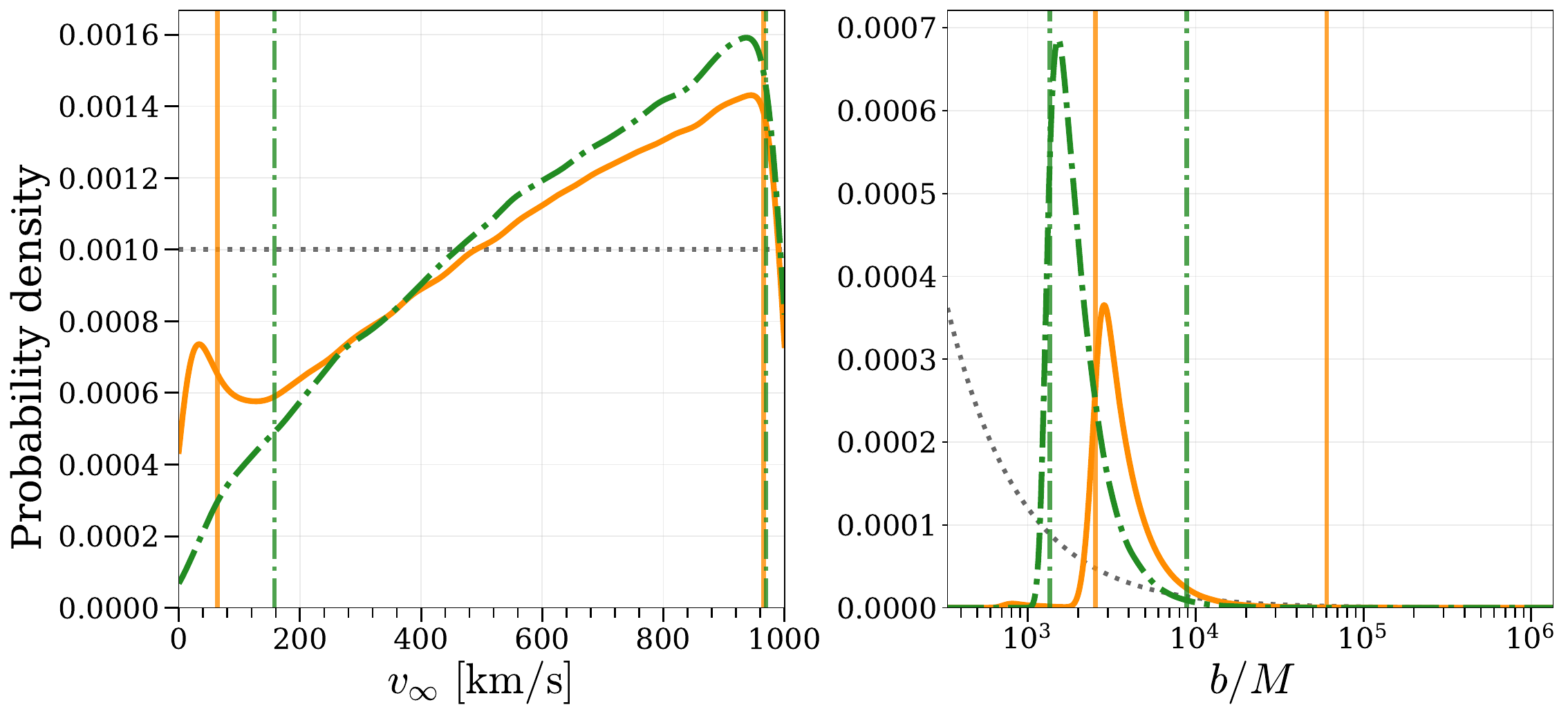}
  \caption{\label{fig:combined posteriors 2}Inferred posteriors for $v_\infty$ and $b$ for GW190521 using eccentric posteriors from \citet{RomeroShaw2020GW190521} (orange/solid) and \citet{Gayathri2020NR} (green/dashed-dotted). Quantities are the same as in Figure \ref{fig:combined posteriors}.}
\end{figure*}

\begin{figure}[t]
    \centering
    \includegraphics[width=0.5\columnwidth]{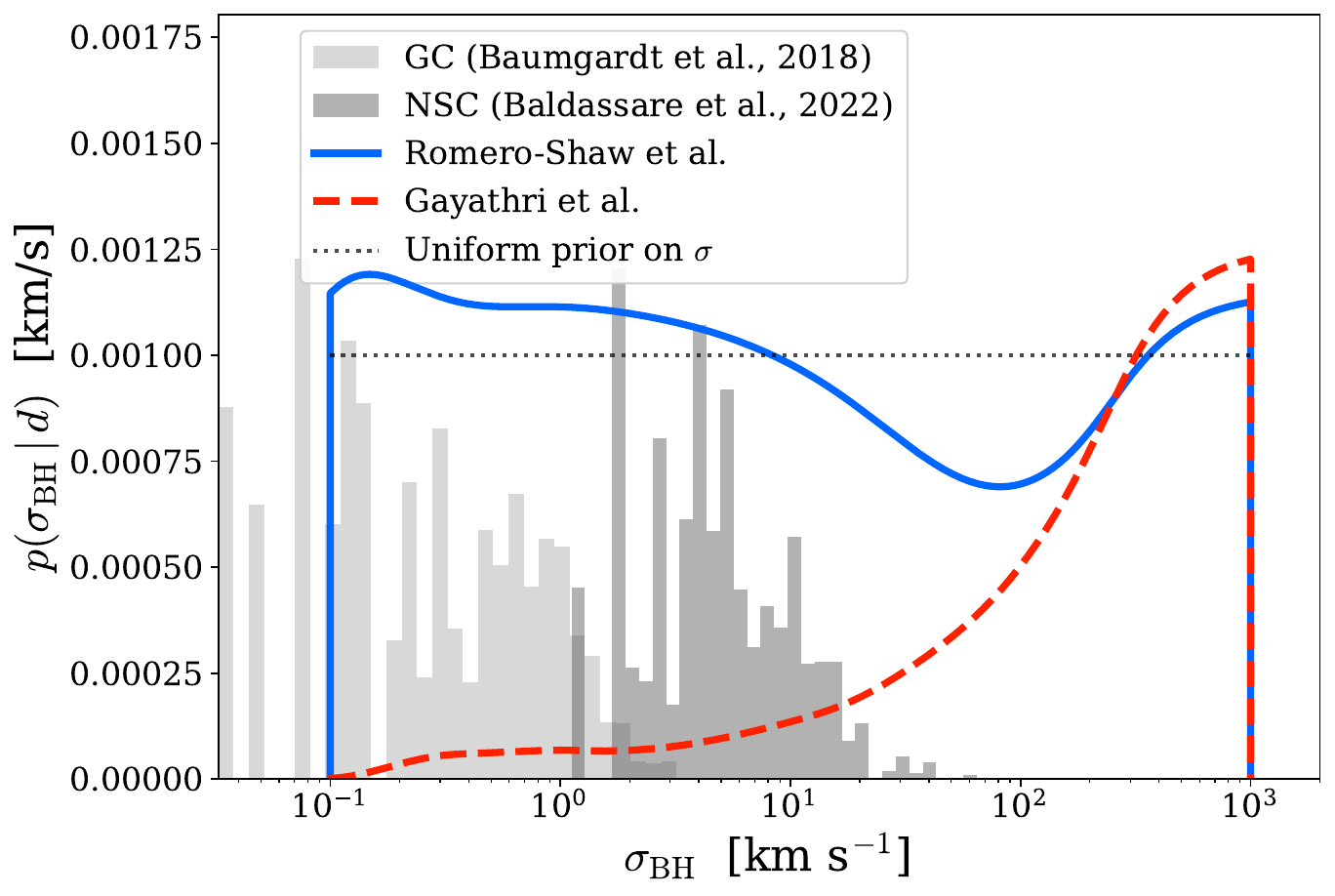}
    \caption{\label{fig:sigmaposteriors-GW190521}  Probability distribution of $\sigma_{BH}$ for the host environment of GW190521 obtained using the formalism in Sec.~\ref{subsec:sigma_single_event} with $\sigma_i=\sigma_{\rm BH}$ (red and blue curves). Gray histograms show observational constraints on $\sigma_{BH}$ based on rescaling stellar velocity dispersions via the equipartition theorem (see text for details). The relative heights of the GC and NSC distributions are also rescaled for better visualization.} 
\end{figure}

Figure~\ref{fig:combined posteriors 2} shows that the inferred relative velocities are large, even in the moderate eccentricity case ($e_{10} \sim 0.1$; \citet{RomeroShaw2020GW190521}). However, a bi-modality arises because of non-negligible low eccentricity support in the posterior. The inferred $v_\infty$ posterior extends to significantly higher values, reaching $\sim10^3\, \mathrm{km/s}$. This indicates that GW190521 requires substantially higher relative velocities compared to GW200105. Figure \ref{fig:sigmaposteriors-GW190521} shows the constraints on the host environment velocity dispersion assuming a capture scenario for GW190521. Since this is a BBH merger, the velocity dispersion here refers to $\sigma_{\rm BH}$. The relative velocity dispersion in Eq.~\eqref{Maxwell_dist} was taken to be $\sigma_{\rm rel}^2 = 2 \sigma_{\rm BH}^2$. The observational stellar velocity dispersion histograms from \citet{Baumgardt_velocity_disp_GC} for GCs and \citet{Baldassare2022MassiveBHNSC} for NSCs were rescaled using $\sigma_{\star} = \sigma_{\rm BH} (m_{\rm BH}/m_{\star})^{1/2}$ assuming $m_{\star} =0.8 M_{\odot}$ and $m_{\rm BH}=30 M_{\odot}$. The calculation of $p_{\rm env}(\sigma_i)$ in Eq.~\eqref{prob cluster} relies on log-normal fits to these histograms. The relevant fit parameters for both the GW190521 and the GW200105 case (treated in the main text) are given in Table \ref{tab:lognormal_fits}.  Using the eccentricity posteriors from~\citet{RomeroShaw2020GW190521}, we estimate that GCs (NSCs) have a 54\% (46\%) probability of hosting GW190521. From \citet{Gayathri2020NR}, we obtain a 31\% (69\%) probability for GCs (NSCs) to be the host.

 \begin{table}[t]
\centering
\caption{ Parameters for log-normal fits to the central compact object velocity dispersion
distributions, $p_{\rm env}(\sigma_i) = \mathcal{LN}(\mu,\, s)$. These are found by rescaling the cluster's central stellar velocity dispersion via the equipartition theorem and assuming all the central stars have mass $m_\star = 0.8\,M_\odot$ while each NS or BH in the subpopulation is assumed to have identical mass $m_c$. The stellar velocity dispersions are taken from Table 2 of \citet{Baumgardt_velocity_disp_GC} for GCs and Table 1 of \citet{Baldassare2022MassiveBHNSC} for NSCs. For central compact object velocity dispersion $\sigma_i$ ($i=$BH or NS), $\sigma_{\rm median}$ is the median value across all clusters in the sample, $\mu=\ln(\sigma_{\rm median})$ is the log of that value, and $s$ is the standard deviation
of $\ln\sigma_i$ across the cluster sample.}
\label{tab:lognormal_fits}
\begin{tabular}{llccccr}
\hline\hline
Event & Environment & $m_c\ [M_\odot]$  & $\mu$ & $s$ &
$\sigma_{\rm median}\ [\rm km/s]$ \\
\hline
\multirow{2}{*}{GW200105} & GC  & \multirow{2}{*}{1.4}  &  $1.144$ & $0.875$ &  $3.14$ \\
                          & NSC &                       &  $3.635$ & $0.786$& $37.92$ \\
\hline
\multirow{2}{*}{GW190521} & GC  & \multirow{2}{*}{30.0} &0.388& $0.875$ &  $0.68$ \\
                          & NSC &                       & $2.103$ &$0.786$ &  $8.19$ \\
\hline\hline
\end{tabular}
\end{table}

\section{Decay time}\label{app:decaytime}
Our analysis also allows us to infer the decay time from the formation of a bound binary to the time the binary enters the reference frequency of a ground-based detector network (taken here to be $20\,{\rm Hz}$). This is straightforwardly computed in our framework by feeding samples that relate the initial and final semi-latus rectum and eccentricity $(x_i, e_i, x_f, e_f)$ (see Sec.~\ref{sec:ecc-scatter}) into Eq.~(3.9) of \citet{TuckerandWill2021}. [Alternatively, one could use Eq.~5.14 of \citet{Peters1964}.] Because the time to merge from $f\sim 10\,{\rm Hz}$ is $\ll 1$ day, this is equivalent to the time from capture to binary merger.

\begin{figure*}[t]
\centering
\includegraphics[width=0.4\linewidth]{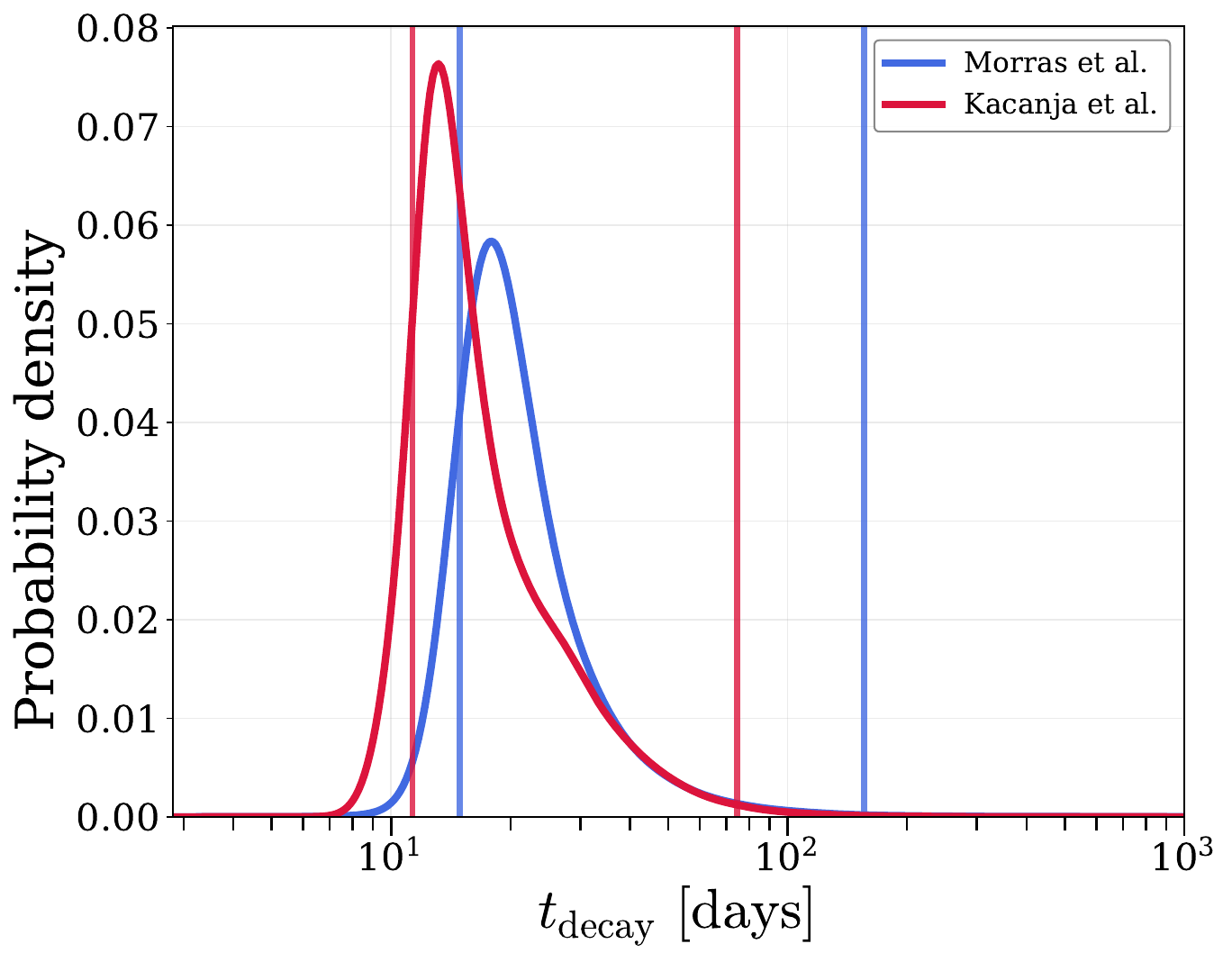}
  \caption{Probability distribution of the binary's decay time for GW200105, the time from capture until the binary reaches the detector frequency band at $f=20\, {\rm Hz}$. The calculation makes use of the eccentric posteriors from either \citet{Morras2025} (blue/solid) or \citet{KacanjaEcc} (red/dashed). Decay times are $15\mbox{--}156$ days in the case of~\citet{Morras2025} and $11\mbox{--}75$ days in case of~\citet{KacanjaEcc} with 90\% credibility.}
  \label{fig:decay time}
\end{figure*}

Figure \ref{fig:decay time} shows the resulting probability distribution for the decay time for GW200105. There, we see that the decay time is of order a few tens to hundreds of days, which is very short compared to astrophysical timescales. This is the first LVK event that allows us to compute the decay time, assuming the binary was formed via GW capture. (Due to the large uncertainty in the final eccentricity inferences for GW190521, we did not consider the decay time calculation for that event.)  

This has several implications. First, due to the very short inferred binary lifetime for this event, we can ignore cluster evolution during the timespan from capture to binary merger. This implies that the velocity dispersion at binary formation and merger is the same. Second, there is the possibility that the remnant of GW200105 may pair up with another black hole or neutron star, depending on the compact object density in the local environment. If that region has a high compact object density, it is also possible that the black hole primary in GW200105 was the product of a previous binary merger. Future observations of similar systems may shed light on this possibility.

     
\bibliographystyle{aasjournalv7}
\bibliography{references}
\end{document}